\long\def\symbolfootnote[#1]#2{\begingroup%
\def\thefootnote{\fnsymbol{footnote}}\footnote[#1]{#2}\endgroup} 

\def\limepy{\textsc{\small{limepy}}}
\def\emcee{\textsc{\small{emcee}}}

\def\python{\textsc{\small{python}}}

\def\intfrac[#1,#2]{I_{#1}^{#2}}
\def\derfrac[#1,#2]{D_{#1}^{#2}}

\def\Eg{E_\gamma}

\def\rh{r_{\rm h}}

\def\ra{r_{\rm a}}

\def\rt{r_{\rm t}}

\def\phit{\phit_{\rm t}}

\documentclass[useAMS,usenatbib, times]{mn2e}
\usepackage{amssymb,latexsym,graphicx,natbib,eufrak,times,amsmath,xfrac,nicefrac}
\usepackage[caption=false]{subfig}

\title{Radial anisotropy in $\omega$ Cen limiting the room for an intermediate-mass black hole}
\author[Alice Zocchi, Mark Gieles, Vincent H\'enault-Brunet]
  {Alice Zocchi$^{1,2}$\thanks{E-mail:
alice.zocchi2@unibo.it}, Mark Gieles$^{1}$, Vincent H\'enault-Brunet$^{1,3}$ \\
$^1$ Department of Physics, University of Surrey, Guildford, GU2 7XH,UK. \\
$^2$ Dipartimento di Fisica e Astronomia, Universit\`{a} degli Studi di Bologna, viale Berti Pichat 6/2, I–40127, Bologna, Italy. \\
$^3$ Department of Astrophysics/IMAPP, Radboud University, PO Box 9010, NL-6500 GL Nijmegen, the Netherlands. \\
}

\date{Accepted ? ?. Received ? ?; in original form ? ?}

\def\LaTeX{L\kern-.36em\raise.3ex\hbox{a}\kern-.15em
    T\kern-.1667em\lower.7ex\hbox{E}\kern-.125emX}

\begin{document}         
\maketitle
\begin{abstract}
Finding an intermediate-mass black hole (IMBH) in a globular cluster (or proving its absence) would provide valuable insights into our understanding of galaxy formation and evolution. However, it is challenging to identify a unique signature of an IMBH that cannot be accounted for by other processes. Observational claims of IMBH detection are indeed often based on analyses of the kinematics of stars in the cluster core, the most common signature being a rise in the velocity dispersion profile towards the centre of the system. Unfortunately, this IMBH signal is degenerate with the presence of radially-biased pressure anisotropy in the globular cluster. To explore the role of anisotropy in shaping the observational kinematics of clusters, we analyse the case of $\omega$ Cen by comparing the observed profiles to those calculated from the family of \limepy\ models, that account for the presence of anisotropy in the system in a physically motivated way. The best-fit radially anisotropic models reproduce the observational profiles well, and describe the central kinematics as derived from Hubble Space Telescope proper motions without the need for an IMBH.
\end{abstract}

\section{Introduction}
\label{Sect_Introduction}

The globular cluster $\omega$ Cen (NGC 5139) is peculiar and extreme in many aspects. It is one of the most massive \citep{Meylan1995,vandeVen2006} and less relaxed \citep{Meylan1987,Meylan1995,GielesHeggieZhao2011} in our Galaxy, showing no strong sign of mass segregation \citep{Anderson2002,Miocchi2010,Goldsbury2013}. It rotates \citep{MeylanMayor1986,Norris1997,Merritt1997,vanLeeuwen2002,Pancino2007} and it is flattened \citep{Geyer1983,WhiteShawl1987}. It is known to host multiple stellar populations, having different chemical abundances, identified as distinct main sequences \citep[$\omega$ Cen was the first cluster for which this feature was observed,][]{Anderson1997,Anderson2002} and subgiant branches \citep{Lee1999,Pancino2000,Ferraro2004,Sollima2005b,Bellini2009}, that appear to be spatially separated \citep{Norris1997,Pancino2003,Pancino2007} and to have a spread in age and metallicity which is not found in most other Galactic globular clusters \citep{Villanova2014}. Its many peculiarities have triggered many observational studies, making it the star cluster with the largest available set of both photometric and kinematic data. The dynamics of $\omega$ Cen has been investigated by means of a number of different techniques and methods \citep[see for example][among many others]{vandeVen2006,Sollima2009,Jalali2012,BVBZ2013,Watkins2013}, with the aim of understanding its origin and evolution, but still many unanswered questions and puzzles remain. In particular, this cluster has been at the centre of a controversy concerning the presence of an intermediate-mass black hole (IMBH) in its centre. 

Recently, the search for IMBHs in globular clusters has received a lot of interest because finding them could provide useful insights to understand galaxy formation and evolution. Indeed, theories on the formation mechanisms of the super-massive black holes found in the centres of many galaxies \citep{HaringRix2004,vandenBosch2012,Davis2013} suggest that IMBHs could be the seeds for the growth of these objects, and it has been proposed that IMBHs could be the result of runaway stellar collisions in dense young globular clusters \citep{PortegiesZwart2004}. Determining if an IMBH is present in a cluster is also useful because it could provide information on the accretion rates of these objects. \citet{Haggard2013} found an upper limit of $3.7-5.3 \times 10^3$ M$_{\sun}$ for the mass of a possible IMBH, by measuring the X-ray luminosity in the centre of $\omega$ Cen, considering the expected accretion efficiency. A similar estimate was obtained by \cite{LuKong2011}, who considered deep radio continuum observations of $\omega$ Cen, and concluded that the upper limit of the mass of a possible IMBH is $1.1-5.3 \times 10^3$ M$_{\sun}$; a more conservative estimate by \citet{Strader2012} brings this upper limit down to 870 M$_{\sun}$, based on these radio observations. A more massive IMBH could be present, if its accretion efficiency is much smaller than expected.

Because globular clusters are almost gas-free, the absence of accreting IMBH is no evidence for the absence of an IMBH. The focus has therefore been mostly on finding signatures in the phase-space distribution of the stars. Unfortunately, this is challenging: their presence produces observable signatures, such as the well-known Bahcall \& Wolf density cusp \citep{BahcallWolf1976}, but similar signatures could also be accounted for by other processes. In practice, two signatures are the most studied, and are the basis for the claims of the detection of these objects in Galactic clusters: a shallow cusp in the surface brightness profile \citep{Noyola2006,Noyola2008} and a rise in the velocity dispersion profile towards the centre of the cluster \citep{AvdM2010,Luetz2013,Feldmeier2013}. However, almost identical fingerprints can also be accounted for by other processes: first, it has been shown \citep{Vesperini2010} that mass segregation, core collapse, or the presence of binary stars in the centre also produce a shallow cusp in the brightness profile; secondly, the central rise in the velocity dispersion profile could also be due to the presence of radially-biased pressure anisotropy in the system, which is the topic of this study.

The controversy regarding $\omega$ Cen represents an illustrative example of the kinematic degeneracy just mentioned. The first claim of the presence of an IMBH in this cluster was by \citet{Noyola2008}, who considered a surface brightness profile and a line-of-sight velocity dispersion profile calculated by integrated spectra in the centre of $\omega$ Cen. By performing an analysis based on the Jeans equation, they fitted spherical, isotropic models with the addition of a central IMBH to their data, and estimated that the IMBH mass required to match the observed rise in the central velocity dispersion is $4.0 \times 10^4$ M$_{\odot}$. \citet{AvdM2010} measured proper motions in the innermost part of the cluster, and analysed these data by means of spherical models characterised by radial anisotropy in the core and tangential anisotropy in the outer parts \citep{vdMA2010}. They found an upper limit of $1.2 \times 10^4$ M$_{\odot}$ for the central IMBH mass, and they showed that when considering cored models no IMBH is needed to reproduce the data. Recently, by comparing the results of numerical simulations with data of Galactic globular clusters, \citet{Baumgardt2017} found that $\omega$ Cen is the only cluster for which the presence of a central IMBH with mass $\sim 4 \times 10^4$ M$_{\odot}$ is needed in order to reproduce the central rise observed in the velocity dispersion profile. 

The discrepancy in these estimates could be due to several reasons: these works use different data sets, models with different amounts of anisotropy, and a different position of the centre to calculate the radial profiles they considered. \citet{Noyola2010} explored the effects introduced by the last one: they produced different velocity dispersion profiles by means of integrated spectra obtained in the central region of $\omega$ Cen, using different assumptions for the position of the centre of the cluster, and then compared them with their spherical and isotropic models, finding that the best-fit black hole mass ranges from $3 \times 10^4$ M$_{\odot}$ to $5.2 \times 10^4$ M$_{\odot}$, depending on the choice of the centre.

This paper focuses on the second possible source of the discrepancy among those outlined above: we explore the role of anisotropy in determining the shape of the observable profiles of $\omega$ Cen. We will do this by using a family of dynamical models defined from a distribution function \citep{GielesZocchi}, referred to as the \limepy\ models, which enable the calculation of isotropic models as well as models containing different amounts of radial anisotropy. These models describe systems that are isotropic in the centre, radially anisotropic in the intermediate part and isotropic again at the truncation (this shape of anisotropy profile is physically motivated: we refer to Section~\ref{Sect_Models_Anis} for a detailed explanation). In the past, some concerns have been raised regarding the fact that the models used by \citet{vdMA2010} to describe $\omega$ Cen require the cluster to be radially anisotropic in the core. Indeed, \citet{Tiongco2016} recently showed that the innermost parts of globular clusters isotropize quickly, and that radial anisotropy is expected to be found in their intermediate parts. \limepy\ models are isotropic in the centre, but despite this, the anisotropy in the outer parts results in an increase of the line-of-sight velocity dispersion in the centre of the cluster (in projection), as we will show in Section~\ref{Sect_VD_fit}.

In order to study the importance of anisotropy in shaping the observable profiles, we consider different sub-families of \limepy\ models, each one characterised by a well-defined global amount of anisotropy, and then we fit them to the data with a two-step procedure. First, we use the surface brightness profile to determine the best-fit model parameters. Then, we compare these models to the line-of-sight velocity dispersion profile to obtain a velocity scale, that translates into a measure of the mass of the cluster. Only after all the parameters have been determined, we compare the models to proper motions measurements. This procedure, in a similar way as the analysis based on the Jeans equation carried out by \citet{Noyola2010} and \citet{vdMA2010}, ensures that the shapes of the radial profiles of all the considered quantities are set by imposing that the models accurately describe the surface brightness profile, thereby isolating the effect of varying the anisotropy on the kinematics (we note that varying the anisotropy parameter in \limepy\ models affects both the spatial density and the kinematics). To understand which flavour of anisotropy is to be preferred when describing $\omega$ Cen, we also carry out an additional one-step fit, by considering all the data profiles, thus determining all the parameters at the same time.

The paper is organised as follows. In Section~\ref{Sect_Models} we introduce the models we used to carry out our analysis, and Section~\ref{Sect_Data} describes the data we considered. In Section~\ref{Sect_Methods} we present the results of the two-step and of the one-step fitting procedure. In Section~\ref{Sect_Results} we discuss the results by comparing them to those from previous works; the implications for the presence of an IMBH are also outlined. Finally, we present our conclusions in Section~\ref{Sect_Conclusion}.

\section{Models}
\label{Sect_Models}

To explore the effect of a certain amount of radial anisotropy on the observed properties of the cluster, we consider the family of \limepy\ models\footnote{The \limepy\ (Lowered Isothermal Model Explorer in PYthon) code is available from https://github.com/mgieles/limepy.} introduced by \citet{GielesZocchi}. For these self-consistent models, it is possible to compute several quantities such as the projected density profile and the projected velocity dispersion components (along the line of sight and on the plane of the sky), to then compare them with the data. 

\subsection{Distribution function}

This family of models is defined by a distribution function, depending on the specific energy $E$ and angular momentum $J$:
\begin{eqnarray}
f(E,J^2)\! = \!
\begin{cases}
A \exp \left(\displaystyle\!-\!\frac{J^2}{2 \ra^2 s^2} \!\right) \! \Eg \left(\! \displaystyle g, \frac{\phi(\rt)\!-\!E}{s^2} \!\right) & E\!<\!\phi(\rt) \\
0 & E\!\geq\!\phi(\rt) \ ,
\end{cases}
\label{Eq_DF_Limepy}
\end{eqnarray}
where $\phi$ is the potential, and $\rt$ is the truncation radius. The function $\Eg(a, x)$ in equation~(\ref{Eq_DF_Limepy}) is defined as
\begin{align}
\Eg(a, x) =  
\begin{cases}
\exp(x)  &  a=0 \\
\displaystyle\exp(x) \frac{\gamma(a, x)}{\Gamma(a)}   &  a>0 \ ,
\end{cases}
\label{eq:eg}
\end{align}
where $\gamma(a, x)$ is the lower incomplete gamma function, and $\Gamma(a)$ is the gamma function. The expression for the density of these models can be written by means of special functions that depend on the potential $\phi$ and on $r$, making the Poisson equation fast to solve\footnote{Having an expression of the density $\rho$ as a function of $\phi$ and on $r$ prevents us from calculating it numerically, by solving an integral at each considered radial position. This speeds up the solution of the Poisson equation significantly.}, and the potential and all the other derived quantities of the models straightforward to calculate.

Starting from the distribution function, several quantities can be computed. For example, it is straightforward to compute the density and all the components of the velocity dispersion for the models. By projecting these quantities along the line of sight, it is possible to obtain the surface density, and the line-of-sight and proper motions component of the velocity dispersion to be compared with the data. Because these models describe a homogeneous population\footnote{A definition of \limepy\ models with multiple mass components is available; for an example of application, see \citet{Peuten2016}.}, i.e. it can be represented as a system of stars with the same mass and the same luminosity, it is possible to transform the mass density into luminosity density to be compared with the surface brightness data by assuming a constant mass-to-light ratio over the entire extent of the cluster. This is an approximation that holds in the case of clusters that are not much mass-segregated \citep{Anderson2002,Miocchi2010,Goldsbury2013}. The effect of mass segregation, indeed, causes stars of different mass to have a different spatial distribution in the system and different kinematics (the less massive stars being less centrally concentrated and having a larger central velocity dispersion, for example); we will explore this issue in this context in a follow-up paper.

\subsection{Model parameters}

Each model in the family is identified by specifying the values of three parameters. The parameter $W_0$ is the central dimensionless potential, and is also referred to as the \textit{concentration parameter}. It is used as one of the boundary conditions in solving the Poisson equation, and determines the shape of the radial profiles of some relevant quantities. This parameter has the same role as in, for example, \citet{King1966} and \citet{Wilson1975} models. The \textit{anisotropy radius} $\ra$ sets the amount of anisotropy contained in the system. The smaller it is, the more radially anisotropic is the model; when $\ra$ is large with respect to the truncation radius, the model is everywhere isotropic. The anisotropy is included in the models in the same way as in \citet{Michie1963} models.

The third parameter, that differentiates \limepy\ models from the ones mentioned above, is the \textit{truncation parameter} $g$, which sets the sharpness of the truncation in energy \citep[see also][]{G-L_V2014}: the larger it is, the more extended the models, and the less abrupt the truncation. To further clarify its role, we point out that for $\ra \rightarrow \infty$ and by setting $g = 0, 1,$ and $2$, we obtain the \citet{Woolley1954}, \citet{King1966}, and (isotropic, non-rotating) \citet{Wilson1975} models, respectively. The choice of this family of models, therefore, is also supported by the fact that it includes these well-known models, that have been widely used to describe Galactic globular clusters.

In addition, it is necessary to define the velocity, radial, and mass scales, in order to describe every property of the model in the desired set of units. These are related to $A$ and $s$, which represent a mass density in 6-dimensional phase space and a velocity scale, thus naturally defining a radial scale via the gravitational constant $G$. We note that it is enough to specify two of the three scales to completely determine the set of units to use: for this reason, we decided to consider the mass scale and radial scale as fitting parameters, as these are intuitive when fitting the models to data.

\subsection{Radial anisotropy}
\label{Sect_Models_Anis}

The \limepy\ models describe spherically-symmetric systems, therefore their main properties can be expressed as a function of the distance from the centre, $r$. When considering anisotropic configurations, the models are isotropic in the centre, radially anisotropic in the intermediate part, and isotropic again near the tidal radius $\rt$. 
This particular form of the anisotropy profile makes them adequate to describe globular clusters, because this is expected for the dynamical evolution of stellar systems in an external tidal field. On the one hand, this is indeed obtained when taking into account the effect of tides on the result of the process of collisionless collapse \citep{vanAlbada1982,TrentiBvA2005}: the anisotropy profile of a stellar system generated this way is isotropic in the centre and radially anisotropic in the outer parts, and the interaction with the external tidal field induces isotropy also near the truncation. On the other hand, a similar form of anisotropy is obtained also when considering the collisional collapse of a compact, isotropic, and initially tidally-underfilling globular cluster \citep{Tiongco2016}: during core collapse the stars are scattered outside the core on radial orbits, while the cluster expands; then, when the system fills the Roche lobe, the interaction with the external tidal field tends to isotropise the orbits, as stars gain angular momentum and the amount of radial anisotropy near the truncation is suppressed \citep{OhLin1992}. By analysing several snapshots of a numerical simulation we recently showed that the \limepy\ models accurately reproduce the main properties of such systems during their evolution \citep{Zocchi2016}. 

A global measure of the amount of anisotropy in the system is given by the quantity $\kappa = 2K_{\rm r}/K_{\rm t}$, defined as the ratio of the radial and tangential components of the kinetic energy. Clearly, $\kappa > 1$ for radially anisotropic systems, $\kappa < 1$ for tangentially anisotropic systems, and $\kappa = 1$ for isotropic systems. All the \limepy\ models that we will refer to in this paper are stable against radial orbit instability, as they satisfy the condition $\kappa<1.7\pm0.25$ \citep{PolyachenkoShukhman1981}. A local indication of the anisotropy content of a system is given by the ratio of the tangential to radial components of the velocity dispersion projected on the plane of the sky, $\sigma_{\rm T}/\sigma_{\rm R}$. Therefore, $\sigma_{\rm T}/\sigma_{\rm R}\,{=}\,1$ indicates isotropy, $\sigma_{\rm T}/\sigma_{\rm R}\,{<}\,1$ radially-biased anisotropy, and $\sigma_{\rm T}/\sigma_{\rm R}\,{>}\,1$ tangentially-biased anisotropy. This quantity is useful for comparison with globular clusters for which proper motion data are available, as this is the only direct observable measure of the anisotropy of a stellar system.

\section{Data}
\label{Sect_Data}

We briefly outline here the data sets used for our analysis, and we explain how we calculated the observational profiles to be compared with the models.

\subsection{Photometric data}

We used the surface brightness profile derived by \citet{TKD1995}. We decided to complement this profile by adding the one derived by \citet{Noyola2008}, providing data on the innermost region. The composite profile consists of 72 data points in total, given by the $V$-band surface brightness $\mu_{V,i}(R_i)$ measured in mag arcsec$^{-2}$ at the (projected) radial position $R_i$ in arcsec.

Before comparing these data with the models, we adopted an extinction correction, assuming a constant extinction over the entire extent of the cluster. We computed $A_V$ by using the value of the reddening given in the \citet{Harris2010} Catalog, and corrected the surface brightness to obtain $\mu_i(R_i) = \mu_{V,i}(R_i) - A_V$ for each data point. We did not apply any other correction to the profile, assuming that any foreground and background contamination had been already removed from the data.

Uncertainties on the data by \citet{Noyola2008} are directly provided, while we calculated those on the data by \citet{TKD1995} by following the procedure adopted by \citet{MLvdM2005}, starting from the weights $w_i$ associated to each data point provided by \citet{TKD1995}. As suggested by \citet{MLvdM2005}, we excluded the points with weights $w_i<0.15$ in the original profile. The uncertainties are computed as $\delta \mu_i = \sigma_{\mu}/w_i$, with the base error bar $\sigma_{\mu} = 0.142$ for $\omega$ Cen \citep[for more details on this procedure and on the meaning of $\sigma_{\mu}$, we refer the reader to Section 2.2 of][]{MLvdM2005}.

To carry out the fit, as suggested by \citet{MLvdM2005}, we transformed the magnitudes in luminosities $l_i$ \citep[for more details, see also Appendix B of][]{ZBV2012}. This ensures that the $\chi^2$ function is more stable, and allows us to make a more straightforward comparison with the models.

\subsection{Kinematic data}
\label{Sect_Data_Kin}

We combined two different data sets of line-of-sight velocities of stars in $\omega$ Cen, the one by \citet{Reijns2006} and the one by \citet{Pancino2007}. After determining which stars are in common in the two samples, we kept only the measure with the smallest associated error. We thus built a combined sample consisting of radial velocities measured for 1868 stars in the cluster, covering a spatial extent from $\sim 11$ arcsec to $\sim 1800$ arcsec from the cluster centre\footnote{In our analysis, we consider the position of the centre of $\omega$ Cen to be the one given by \citet{AvdM2010}.}. For each star, the position on the sky, the line-of-sight velocity, and the associated error are provided. We computed the line-of-sight velocity dispersion profile $\sigma_{\rm los}$ by following the procedure described in \citet{PryorMeylan1993}.

Two data sets of proper motions measurements are available for this cluster. The data set by \citet{vanLeeuwen2000} consists of ground-based measurements of proper motions for 9847 stars, 2740 of which are then considered for the fit \citep[for a more detailed description of the treatment of these data, see Appendix A in][]{BVBZ2013}. The data set by \citet{AvdM2010} provides Hubble Space Telescope (HST) measurements for 72,970 stars. For each star, the position on the sky, the radial and tangential components of the velocity on the plane of the sky, and the associated error are provided. In this work, we considered the radial and tangential proper motion velocity dispersion profiles, $\sigma_{\rm R}$ and $\sigma_{\rm T}$, and the projected anisotropy profile $\sigma_{\rm T}/\sigma_{\rm R}$ as they were calculated by \citet{BVBZ2013}. The ground-based proper motions have large uncertainties, but because of the wide-field coverage they provide, we do include these data in our analysis. We will not be able to place strong constraints on the anisotropy with these data, but we note that with the second Gaia data release (end 2017/early 2018) we will be able to sample the entire anisotropy profile of $\omega$ Cen to address this in greater detail \citep{Pancino2017}.

\section{Fitting procedure and results}
\label{Sect_Methods}

Our goal is to show how the presence of radially-biased anisotropy in a system can alter the observable quantities, and in particular to show its effect on the projected velocity dispersion profile, which is the main observable used to infer the presence of IMBHs in globular clusters. 

\subsection{Two-steps fitting procedure}
\label{Sect_TwoSteps}

Previous works \citep[e.g.][]{vdMA2010,Noyola2010,Watkins2013} analysed the data of $\omega$ Cen by using the Jeans equation to infer the presence of an IMBH, by assuming a certain functional form for the anisotropy of the cluster. The Jeans equations are obtained by taking velocity moments of the collisionless Boltzmann equation. In the case of spherically symmetric systems, there is only one equation for the radial coordinate, which relates the mass density, the radial component of the velocity dispersion, and the anisotropy. This equation is often used to obtain the velocity dispersion of a stellar system from the surface brightness profile properly deprojected by assuming a radial profile for the mass-to-light ratio and for the anisotropy function. The first input needed in this approach is therefore the surface brightness profile. One of the shortcomings of this approach is that it is not a priori guaranteed that there exists a supporting distribution function corresponding to the solutions that are found. Because the anisotropy type and profile are assumed arbitrarily, and manually changed until the obtained velocity dispersion reproduces the data, the modelling is degenerate: different assumptions for the anisotropy profile can produce similar observable features \citep[this is the so called mass-anisotropy degeneracy; see for example][among many others]{BinneyMamon1982,Tonry1983,Lokas2002}.

Here we address this problem from a different point of view, by means of dynamical models defined from a distribution function, with a physically justified anisotropy profile that resembles the ones observed in numerical simulations of both collisionless and collisional systems (see Section~\ref{Sect_Models_Anis}). In order to follow a procedure as close as possible to the one used in analyses based on the Jeans equation (i.e., in order to use the surface brightness profile as a starting point to determine the dynamics of the system), we decided to carry out two steps in our fitting procedure. Because our models include the anisotropy self-consistently, we identify several models in the \limepy\ family by means of the values of the anisotropy parameter $\kappa$, which gives a global indication of the amount of anisotropy in the system. This corresponds to the anisotropy assumption in the Jeans analysis.

The first step in our fitting procedure involves the surface brightness profile, to determine the best-fit model parameters for the chosen amount of anisotropy. This procedure ensures that the structural parameters of the models depend only on the surface brightness profile, which is therefore responsible for determining the shape of all the other profiles we will consider. 

In the second step we find the vertical scaling that is needed to minimise the $\chi^2$ when comparing the line-of-sight velocity dispersion of the model to the data, thus determining the mass-to-light ratio: this is the only parameter depending on the observed kinematics. We point out that the proper motions are not used in any step of this fitting procedure. In Section~\ref{Sect_OneStep} we present a fit to all the data.

\begin{table*}
\begin{center}
\caption[Best-fit parameters for $\omega$ Cen.]{Best-fit parameters for $\omega$ Cen. The first seven rows of the table refer to the two-steps fitting procedure. Each line refers to a different model, identified by the value of the anisotropy parameter $\kappa$ listed in the first column. For each model, several parameters are given, namely the concentration parameter $W_0$, the truncation parameter $g$, the half-mass radius $\rh$ (also given in arcsec in the third column from the right), the truncation radius $\rt$, the mass $M$ and luminosity $L$ of the cluster, the mass-to-light ratio $M/L$, the distance $d$, and the dimensionless anisotropy radius $\hat{r}_{\rm a}$. The last line of the table refers to the fit carried out at once on all the available data, and the same parameters are given also for this best-fit model. The uncertainties are indicated for the best-fit parameters of each model and for the value of the parameter $\kappa$ obtained from the one-step fitting procedure.} 
\label{Tab_Best_Fit_Parameters}
\begin{tabular}{ccccccccccc}
\hline\hline
$\kappa$  & $W_0$ & $g$ & $\rh$ & $\rt$ & $M$ & $L$ & $M/L$ & $\rh$ & $d$ & $\hat{r}_{\rm a}$ \\
 &       &     & [pc]  & [pc]  & [$10^6$ M$_{\odot}$] & [$10^6$ L$_{\odot}$] & [M$_{\odot}$/L$_{\odot}$] & [arcsec] & [kpc] & \\
\hline
1.00 & 4.83 $^{+0.89}_{-1.49}$ & 1.95 $^{+0.57}_{-0.53}$ & 8.74 $\pm$ 0.50 & 101.21 & 2.84 & 1.24 $^{+0.14}_{-0.13}$ & 2.48 $\pm$ 0.03 & 360.68 & 5.00 & $\infty$ \\
1.05 & 4.62 $^{+1.00}_{-1.76}$ & 1.88 $^{+0.65}_{-0.59}$ & 8.65 $^{+0.51}_{-0.46}$ & 102.30 & 2.89 & 1.24 $\pm$ 0.14 & 2.52 $\pm$ 0.03 & 357.00 & 5.00 & 5.30\\
1.10 & 4.48 $^{+0.93}_{-1.45}$ & 1.78 $^{+0.58}_{-0.56}$ & 8.65 $^{+0.45}_{-0.48}$ & 101.84 & 2.94 & 1.24 $^{+0.14}_{-0.13}$  & 2.56 $\pm$ 0.03 & 356.64 & 5.00 & 3.61 \\
1.15 & 4.36 $^{+0.93}_{-1.54}$ & 1.64 $^{+0.61}_{-0.57}$ & 8.62 $^{+0.45}_{-0.49}$ &  95.04 & 2.99 & 1.24 $\pm$ 0.14 & 2.60 $\pm$ 0.03 & 355.80 & 5.00 & 2.87 \\
1.20 & 4.08 $^{+1.02}_{-1.50}$ & 1.60 $^{+0.59}_{-0.56}$ & 8.61 $^{+0.43}_{-0.47}$ &  93.47 & 3.05 & 1.25 $\pm$ 0.13 & 2.63 $\pm$ 0.03 & 355.16 & 5.00 & 2.30 \\
1.25 & 3.88 $^{+1.09}_{-1.62}$ & 1.55 $^{+0.59}_{-0.60}$ & 8.56 $^{+0.44}_{-0.47}$ &  92.57 & 3.08 & 1.24 $\pm$ 0.14 & 2.69 $\pm$ 0.03 & 353.02 & 5.00 & 1.95 \\
1.30 & 3.76 $^{+0.97}_{-1.56}$ & 1.45 $^{+0.57}_{-0.52}$ & 8.60 $^{+0.41}_{-0.45}$ &  87.98 & 3.15 & 1.26 $\pm$ 0.14 & 2.70 $\pm$ 0.03 & 354.61 & 5.00 & 1.74 \\
\hline
1.21 $\pm$ 0.12 & 3.95 $^{+0.86}_{-1.27}$  & 1.82 $^{+0.55}_{-0.61}$  & 8.40 $^{+0.64}_{-0.70}$  & 128.45 & 3.24 $^{+0.51}_{-0.47}$  & 1.20 & 2.92 $^{+0.36}_{-0.32}$  & 346.69 & 5.13 $\pm$ 0.25 & 2.16 $^{+8.36}_{-1.19}$\\
\hline
\end{tabular}
\end{center}
\end{table*}

\subsubsection{Surface brightness profile fitting}
\label{Sect_SB_fit}

As anticipated above, the first step of our procedure consists of fitting the models to the surface brightness profile of $\omega$ Cen, by means of \emcee\ \citep{emcee_paper}, a \python\ implementation of Goodman \& Weare’s affine invariant Markov chain Monte Carlo
ensemble sampler.

When using only the surface brightness profile in the fitting procedure, it is not possible to constrain the amount of anisotropy needed to describe the dynamics of a globular cluster  \citep[this was also shown by][]{ZBV2012}. This is even more so for the \limepy\ family of models, because of the degeneracy between the truncation parameter $g$ and the anisotropy radius $\ra$: both these parameters have a role in determining the extent of the system, so that when we ignore the kinematics, it is not possible to determine if the system is very anisotropic (small $\ra$) and has a sharp truncation (small $g$), or if it is less anisotropic (large $\ra$) and has a shallow truncation (large $g$). Therefore, we carry out the fit seven times, each time by setting $\kappa$ to a different value.

The values of $\kappa$ to use have been chosen by taking into account the results of \citet{TrentiBvA2005}, who showed that for systems originated as a product of collisionless collapse $\kappa \lesssim 1.5$, and of \citet{Zocchi2016}, where the values of $\kappa$ found for a collisional system range from 1.0 to 1.3 during its evolution. However, for very anisotropic models ($\kappa > 1.3$), not all the values of $g$ and $W_0$ are accessible, and this limits range of $\kappa$ that we can consider if we want to reproduce in a satisfactory way the surface brightness profile of $\omega$ Cen. Therefore, we consider an isotropic case along with six radially anisotropic cases, by choosing a sequence of values, equally spaced, in the range $1.0 \leq \kappa \leq 1.3$. This choice of the values of $\kappa$ ensures that all the models are stable against radial orbit instability \citep{PolyachenkoShukhman1981,FridmanPolyachenko1984}.

\begin{figure}
\includegraphics[width=0.48\textwidth]{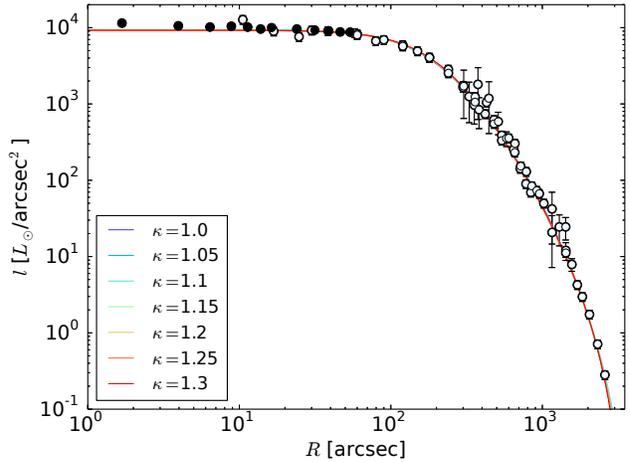}
\caption{Fits by \limepy\ models with different degrees of anisotropy (defined by the value of $\kappa$ as indicated in the legend on the plot) to the surface brightness profile of $\omega$ Cen, here represented in luminosities rather than in magnitudes. Data are from \citet{TKD1995} and \citet{Noyola2008}, and are indicated by open and filled circles, respectively; error bars are shown.}
\label{Fig_SM_SB}
\end{figure}

For the fits, we assume a distance $d = 5.0$ kpc for the cluster, and we determine four free parameters: $W_0$ and $g$, structural model parameters determining the shape of the profile; $L$, the total luminosity of the cluster, in L$_{\odot}$; $\rh$, the half-mass radius of the cluster, in pc. These fitting parameters are found by maximising the log-likelihood function
\begin{equation}
 \Lambda \propto -\frac{\chi^2}{2} \, 
\end{equation}
i.e., by minimising the following quantity
\begin{equation}
 \chi^2 = \sum_{i = 1}^{N_{\rm SB}} \frac{\left[ l_i - \lambda(R_i) \right]^2 }{\delta l_i^2} \ ,
\end{equation}
where $R_i$, $l_i$, and $\delta l_i$ are the radial position, luminosity and luminosity error for each of the $N_{\rm SB}$ points in the surface brightness profile, and the quantity $\lambda$ is the projected luminosity density of the model, its shape depending on the values of all the fitting parameters (this quantity is calculated by scaling the normalised projected density of the model by using the parameters $L$ and $\rh$). For the parameters, we adopt uniform priors over the following ranges: $1 < W_0 < 15$, $0.3 < g < 2.1$, $0.1 < L < 5$ in units of $10^6$ L$_{\odot}$, and $1 < \rh < 20$ in units of pc. The best-fit value for each parameter is identified as the median of the marginalized posterior distribution.

Because $\kappa$ is a quantity that is calculated after solving a model, and not an input model parameter, in order to obtain a model with a certain value of $\kappa$ we carry out an iterative procedure. For each couple of parameters $(W_0,g)$ considered by the walkers of \emcee, we start the iteration by considering a default value for the anisotropy parameter $\ra$. At each step of the iteration, we compare the value of $\kappa$ obtained for the model with the desired one, and we set the new value of $\ra$ accordingly. We stop the iteration when we obtain a model with a $\kappa$ that differs from the target value by less than 0.001.

Figure~\ref{Fig_SM_SB} shows the results of this first step in our fitting procedure. In the plot, the surface brightness profile of $\omega$ Cen is shown (represented in luminosities rather than in magnitude): open circles indicate the points of the profile by \citet{TKD1995}, filled circles those by \citet{Noyola2008}. The best-fit profiles for the seven models with different anisotropy content are also shown, each one indicated with a different colour, as specified by the labels in the figure. It is immediately clear that all the models are able to reproduce the data equally well, and the profiles belonging to the different models are overlapping. This confirms that it is not possible to discriminate between models having a different anisotropy content by only using the surface brightness profile. This also allows us to compare the kinematic profiles of models with a different dynamics that reproduce the surface brightness profile equally well.

The first rows of Table~\ref{Tab_Best_Fit_Parameters} list the values of the best-fit parameters obtained with the fitting procedure described here, together with some other useful quantities. Each line refers to a different model, identified by the value of the anisotropy parameter $\kappa$ listed in the first column. For each model, several best-fit parameters are given: the fitting parameters $W_0$, $g$, $\rh$, $L$, and $M/L$, the derived parameters $\rt$ and $M$, and the assumed value of the distance $d$. In the case of these fits, where we fixed the value of $\kappa$, thus setting the amount of anisotropy in the system, we provide the value of the dimensionless anisotropy radius $\hat{r}_{\rm a}$, defined as the ratio of the anisotropy radius to the scale radius $\ra/r_{\rm s}$ \citep[see Sect. 2.1.2 of][]{GielesZocchi}, needed to get the desired $\kappa$ for the best-fit models; we recall that the value of $\hat{r}_{\rm a}$ is found iteratively for each model, in order to get models with the desired value of $\kappa$. For the fitting parameters, we also show the 1-sigma uncertainties. By inspecting the table, it is clear that the values of the truncation parameter $g$ decrease when the amount of anisotropy in the system increases (increasing $\kappa$, decreasing $\ra$): this is due to the degeneracy described above.

\subsubsection{Line-of-sight velocity dispersion profile fitting}
\label{Sect_VD_fit}

After having found the best-fit model parameters to describe the surface brightness profile for each value of $\kappa$, we compare the kinematic profiles predicted by the models with the observed ones. To do that, we need to find the mass $M$ of the cluster, to get the appropriate velocity scale. Therefore, we obtain the mass-to-light ratio $M/L$ that gives the best-fit to the line-of-sight velocity dispersion profile for each model.

\begin{figure}
\includegraphics[width=0.48\textwidth]{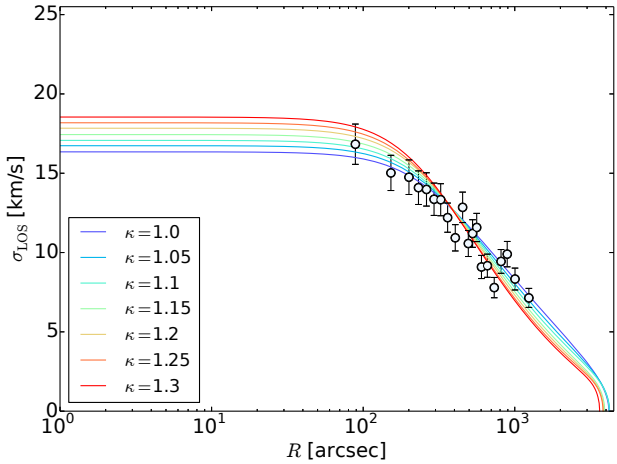}
\caption{Line-of-sight (projected) velocity dispersion profile of $\omega$ Cen, $\sigma_{\rm LOS}$. Data are from \citet{Reijns2006} and \citet{Pancino2007}, with the profile calculated as in \citet{BVBZ2013}; error bars are shown. The lines reproduce the best-fit profiles obtained for the different choices of anisotropy content, as indicated by the labels.}
\label{Fig_SM_VD}
\end{figure}

\begin{figure}
\includegraphics[width=0.48\textwidth]{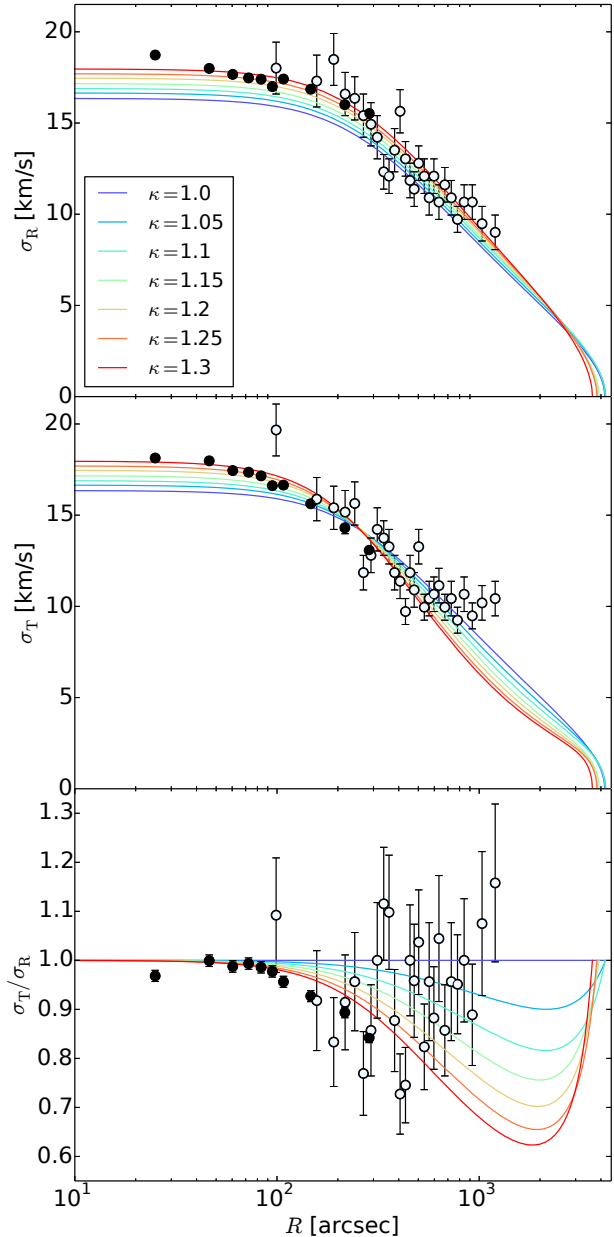}
\caption{Top and middle panels: radial and tangential (proper motions) velocity dispersion profiles for $\omega$ Cen. Bottom: projected anisotropy profile, calculated as the ratio between the tangential and radial components of the velocity dispersion. Data are from \citet{vanLeeuwen2000} and \citet{AvdM2010}, represented with open and filled circles, respectively; error bars are shown. Each line corresponds to a model with a given anisotropy content, as indicated by the labels in the top panel.}
\label{Fig_SM_PM}
\end{figure}

To do this, we minimise the following quantity:
\begin{equation}
 \chi^2 = \sum_{i = 1}^{N_{\rm VD_{los}}} \frac{\left[ \sigma_{{\rm los},i} -  \sigma_{\rm LOS}(R_i) \right]^2 }{\delta \sigma_{{\rm los},i}^2} \ ,
\end{equation}
where $R_i$, $\sigma_{{\rm los},i}$, and $\delta \sigma_{{\rm los},i}$ are the radial position, line-of-sight velocity dispersion and its error for each of the $N_{\rm VD_{los}}$ points in the velocity  dispersion profile, and $\sigma_{\rm LOS}$ is the velocity dispersion profile of the models, projected along the line of sight, and depending on the value of the mass-to-light ratio $M/L$. 

We plot the best-fit velocity dispersion profiles identified in this way in Fig.~\ref{Fig_SM_VD}. The open circles show the line-of-sight velocity dispersion profile calculated from the radial velocity measurements by \citet{Reijns2006} and \citet{Pancino2007}. Each line corresponds to a model with a different amount of radial anisotropy, as indicated in the legend of the plot. 

We notice that the isotropic model predicts the smallest value for the central velocity dispersion. By considering larger amounts of anisotropy, the central value of the velocity dispersion increases. We also note that the profiles of different models have different shapes, contrarily to what we observed for the projected density. The small number of data points and their large error bars (as compared to the ones for the surface brightness profile) suggest that the anisotropy profile can be constrained with more data in the outer parts, in order to cover the entire extent of the cluster.

By inspecting Fig.~\ref{Fig_SM_VD}, it is clear that different models predict different values for the truncation radius $\rt$ of the cluster (see also Table~\ref{Tab_Best_Fit_Parameters}). From Fig.~\ref{Fig_SM_SB} this difference cannot be appreciated, because it occurs beyond the last data point, at very low values of luminosity density. The truncation radius appears to be smaller when increasing the anisotropy (see the discussion about the degeneracy between $g$ and $\ra$ in Sect.~\ref{Sect_SB_fit}).

\subsubsection{Comparison to proper motions data}

After having determined all the model parameters and the scales in the previous two fitting steps, it is then possible to calculate many other properties of the models to compare them with additional data. In particular, here we focus on a comparison with the proper motions data described in Section~\ref{Sect_Data_Kin}.

The top and middle panels of Fig.~\ref{Fig_SM_PM} show the radial and tangential (projected) velocity dispersion profiles of $\omega$ Cen, $\sigma_{\rm R}$ and $\sigma_{\rm T}$. The open circles reproduce the points calculated from the ground-based proper motions measurements by \citet{vanLeeuwen2000}, the filled circles those calculated from HST data by \citet{AvdM2010}. Error bars are also indicated: we notice that the ground-based sample has much larger associated errors than the HST one. Each line corresponds to a model with a different amount of radial anisotropy, as indicated in the legend in the top panel. 

The same behaviour observed for the line-of-sight velocity dispersion is also seen here, with the more anisotropic models providing a larger velocity dispersion in the centre. We notice that, for anisotropic models, the three components of the velocity dispersion considered here have profiles different from one another: this well-known properties of anisotropic systems can be used, with more data covering the entire extent of the cluster, to determine the amount of anisotropy actually present in $\omega$ Cen. The forthcoming Gaia proper motions will be of great importance in this respect.

\begin{figure*}
\includegraphics[width=0.48\textwidth]{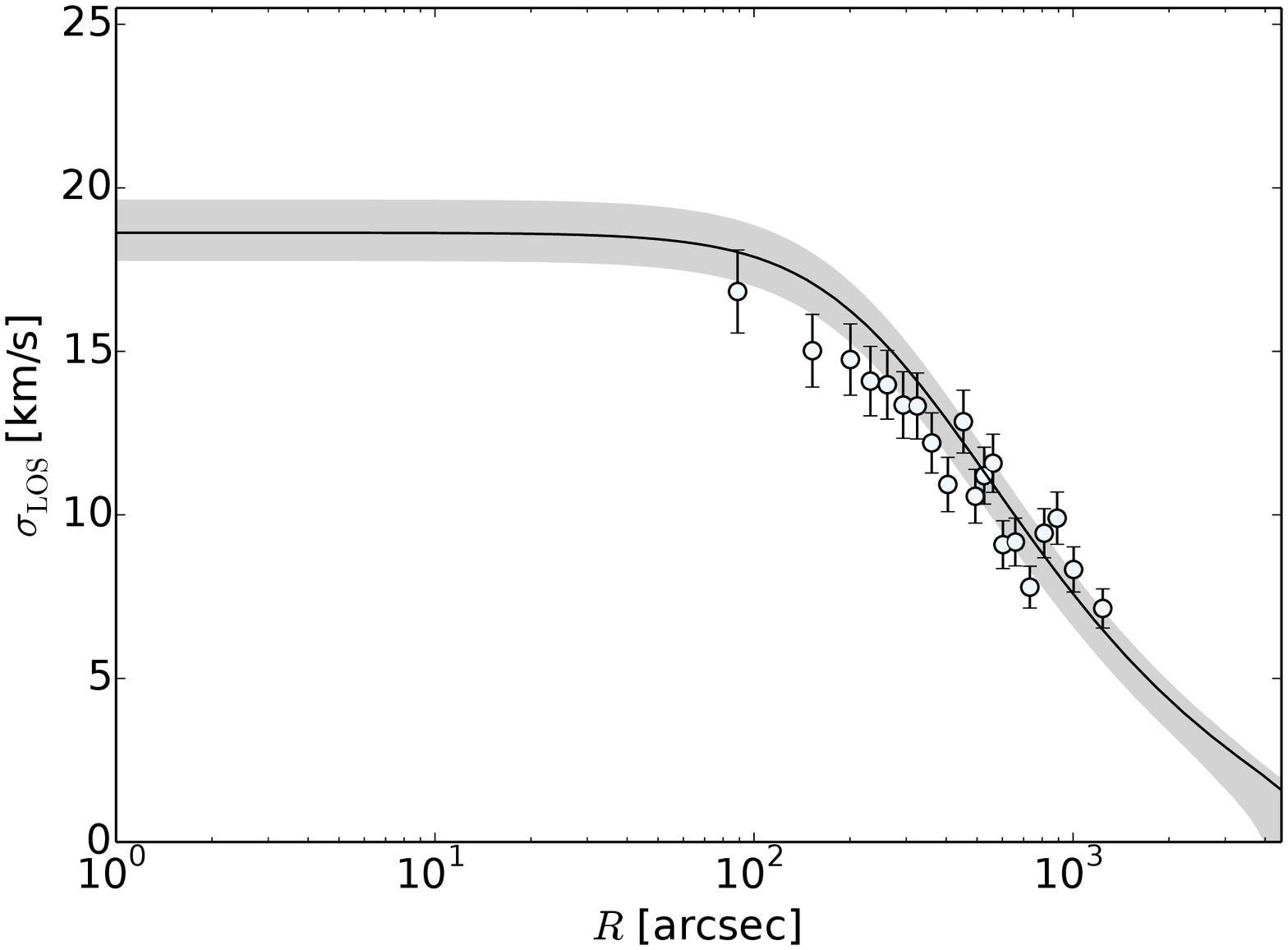}
\includegraphics[width=0.48\textwidth]{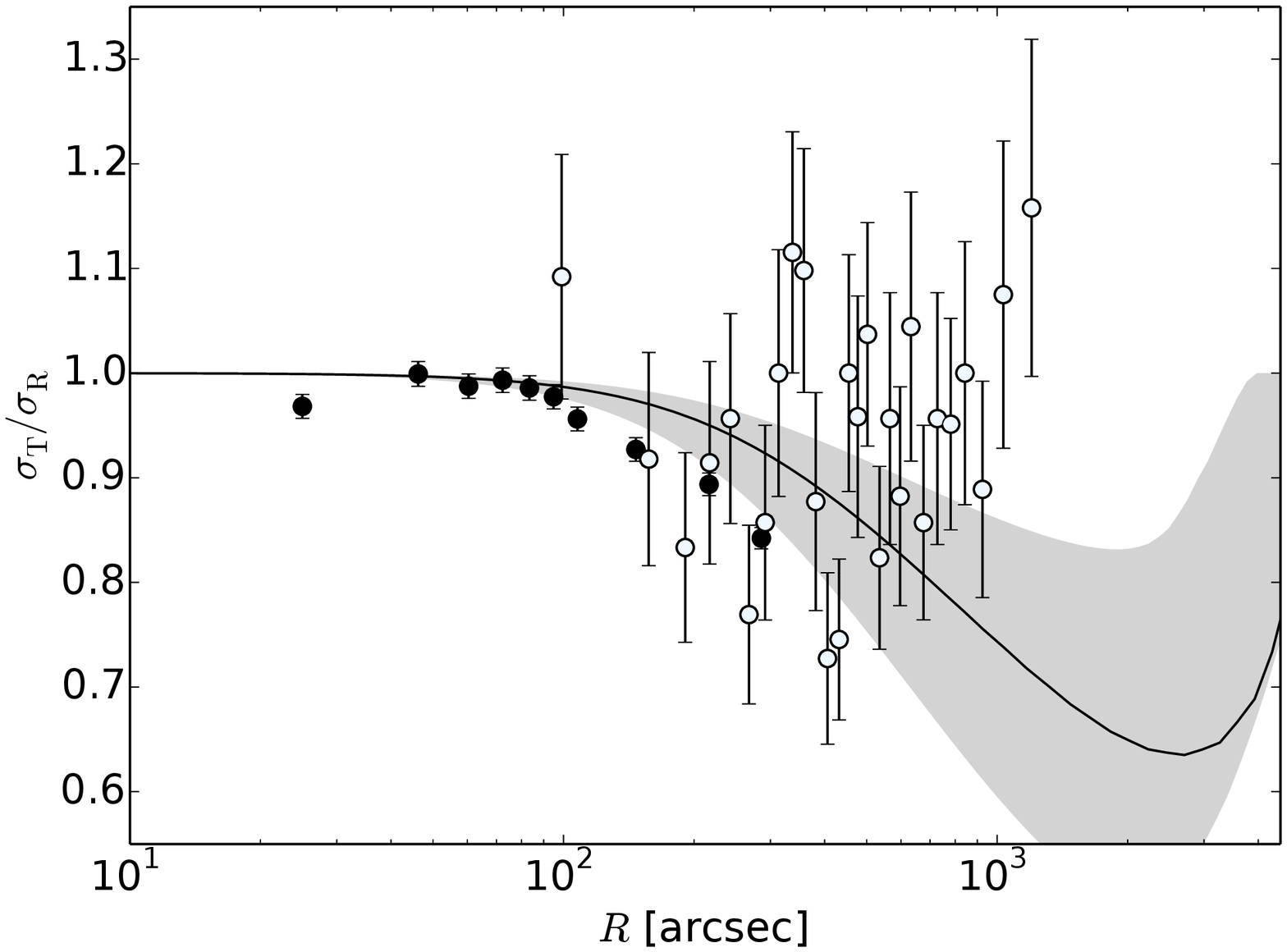} \\
\includegraphics[width=0.48\textwidth]{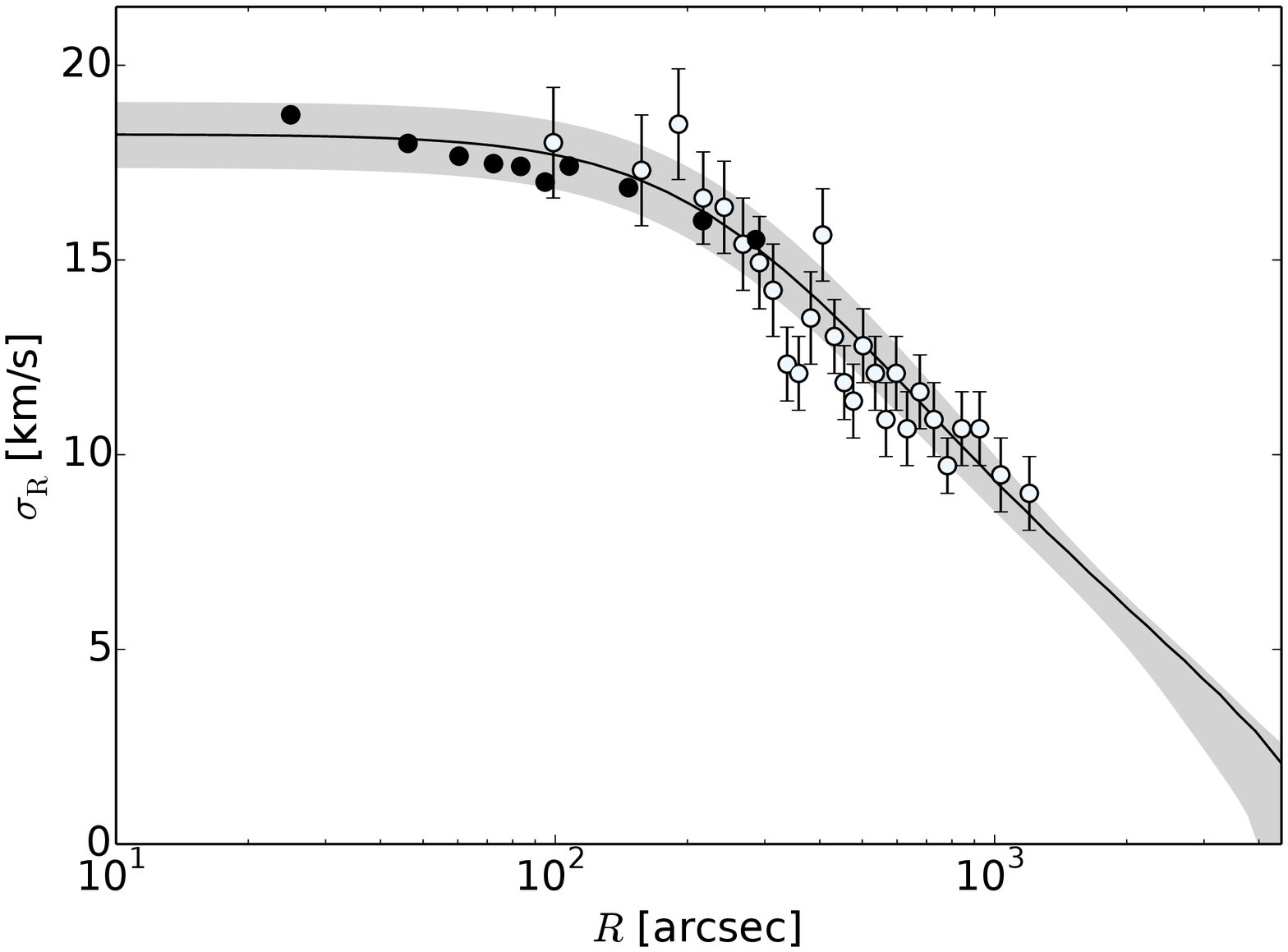}
\includegraphics[width=0.48\textwidth]{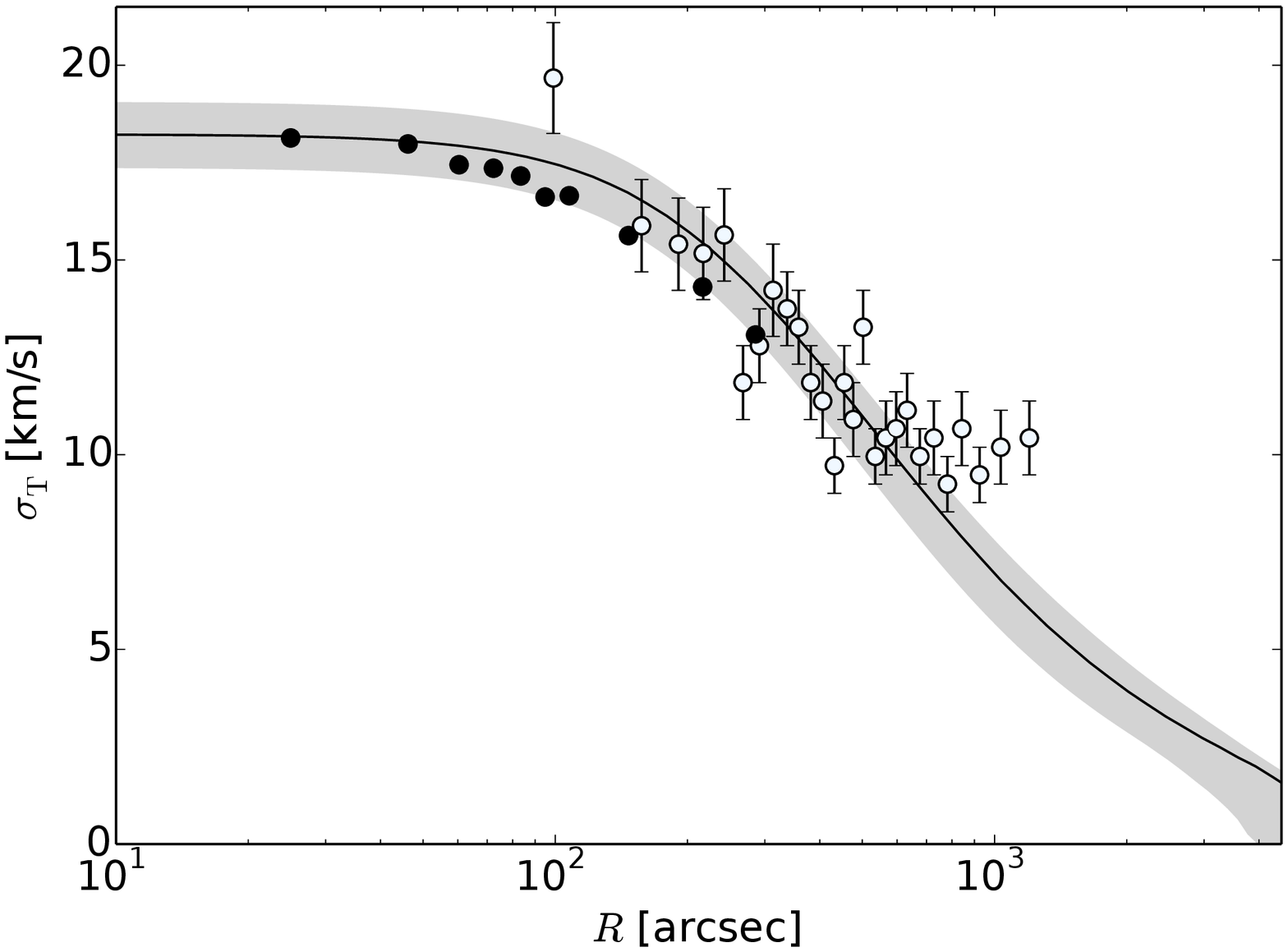}
\caption{Top left: line-of-sight (projected) velocity dispersion profile of $\omega$ Cen, $\sigma_{\rm LOS}$; data are from \citet{Reijns2006} and \citet{Pancino2007}, with the profile calculated as in \citet{BVBZ2013}. Bottom panels: radial and tangential (proper motions) velocity dispersion profiles for $\omega$ Cen. Top right panel: projected anisotropy profile, calculated as the ratio between the tangential and radial components of the velocity dispersion. Proper motions data are from \citet{vanLeeuwen2000} and \citet{AvdM2010}, and are indicated with open and filled circles, respectively. In each panel, the solid line corresponds to the best-fit \limepy\ model obtained by fitting on all the observational profiles at the same time. The shaded grey areas represent the models that occupy a 1-sigma region around the maximum likelihood, as identified by \emcee.}
\label{Fig_All_Kin}
\end{figure*}

The bottom panel of Fig.~\ref{Fig_SM_PM} shows the projected anisotropy profile, calculated as the ratio between the tangential and radial components of the velocity dispersion, $\sigma_{\rm T}/\sigma_{\rm R}$. The HST data seem to indicate a shape of the profile similar to the one reproduced by the most anisotropic model we considered. The ground-based data, instead, have more scatter and larger error bars, and do not follow any clear trend.

We recall that the comparison with proper motions data did not involve any fitting. The fact that the most anisotropic models seem to be a good representation for the observations is an indication that $\omega$ Cen could actually be well described by these anisotropic \limepy\ models. In Section~\ref{Sect_OneStep} we fit the models to all available data.

\subsection{One-step fitting procedure}
\label{Sect_OneStep}

Before comparing the results of our analysis to those in the literature, we present the results of another fit, carried out by considering at the same time all the observational profiles of $\omega$ Cen. Indeed, it is also possible to carry out a unique step in the fitting procedure by considering all the data profiles and by determining all the fitting parameters at once. This procedure, as opposed to the one described in Section~\ref{Sect_TwoSteps}, is also allowing us to fit on the amount of anisotropy present in the system, because here the kinematic data are involved in the fitting procedure. This is interesting because the outcome of this fit gives the best possible representation of the dynamics of $\omega$ Cen by means of the \limepy\ models.

In addition to the structural parameters $W_0$ and $g$, to the mass $M$, half-mass radius $\rh$, and mass-to-light ratio $M/L$ of the system, we also fit on the distance $d$, and on the anisotropy radius $\ra$, which sets the amount of anisotropy of the model. We determine the best-fit values of these 7 parameters by minimising the following quantity:
\begin{equation}
 \chi^2 = \chi_{\rm SB}^2 + \chi_{\rm LOS}^2 + \chi_{\rm PM,R}^2 + \chi_{\rm PM,T}^2
\label{chi_all}
\end{equation}
where each of the terms on the right end side is in the form:
\begin{equation}
 \chi_{\rm X}^2 = \sum_{i = 1}^{N_{\rm X}} \frac{\left[ x_i -  X(R_{i}) \right]^2 }{\delta x_i^2} \ ,
\end{equation}
where $x_i$, $R_i$, and $\delta x_i$ represent the observational quantity, its radial position, and its error, and $X$ is the corresponding model profile, depending on all the fitting parameters. The four terms appearing in equation~(\ref{chi_all}) correspond to the surface brightness profile, line-of-sight velocity dispersion profile, and radial and tangential proper motion velocity dispersion profiles, respectively. For the parameters, we adopt uniform priors over the following ranges: $1 < W_0 < 15$, $0.3 < g < 2.1$, $0.1 < M < 5$ in units of $10^6$ M$_{\odot}$, $1 < \rh < 20$ in units of pc, $1 < d < 10$ in units of kpc, $0.1 < M/L < 5$ in solar units, and $-1 < \log \ra < 2$ (we consider $\log \ra$ as a fitting parameter instead of $\ra$ to have an uninformative prior for this parameter, because it can span several orders of magnitude).

The surface brightness profile of the best-fit \limepy\ model is very similar to the ones presented in Fig.~\ref{Fig_SM_SB}, and it represents the data equally well. Therefore, fitting on all the profiles at the same time has not changed the ability of the models of reproducing this profile.

Figure~\ref{Fig_All_Kin} shows the kinematic profiles of the best-fit \limepy\ model for $\omega$ Cen. Starting from the top left and proceeding counterclockwise, the panels represent the line-of-sight (projected) velocity dispersion profile, the radial and tangential (proper motion) velocity dispersion profiles, and the projected anisotropy profile. With respect to the ones from Section~\ref{Sect_TwoSteps}, these profiles are more extended, and reproduce the proper motions velocity dispersion profiles better. The projected anisotropy profile is still not well matched by the models. This is partly due to the fact that the ground-based data are scattered and have very large errors, and partly due to the fact that the models may not be adequate in representing the internal dynamics of the cluster. Indeed, for example, we know that this cluster is rotating, but the models do not include any rotation.

The best-fit values of the parameters found with this fitting procedure are listed in the last line of Table~\ref{Tab_Best_Fit_Parameters}, together with some other useful quantities. For the fitting parameters we also indicate the uncertainties. We provide the value of $\kappa$ also for this model, as it quantifies the amount of anisotropy in the system, and allows for an immediate comparison with the results of the two-steps fitting results. To make the comparison more meaningful, we also provide the uncertainty obtained for this parameter.


\section{Discussion}
\label{Sect_Results}

In this Section we compare and discuss the results of the analysis we carried out here with those of previous works. 

\subsection{The role of anisotropy}

To further explore the role of anisotropy, we compare the intrinsic $\beta$ profiles, as opposed to the projected ones we used to match the proper motions data. We recall that $\beta$ gives a local measure of the anisotropy in the system, and it is defined as:
\begin{equation}
 \beta = 1 - \frac{\sigma_{\rm t}^2}{2 \sigma_{\rm r}^2}
\end{equation}
where $\sigma_{\rm t}$ and $\sigma_{\rm r}$ are the tangential and radial components of the velocity dispersion. Values of $\beta > 0$ indicate radial anisotropy, $\beta < 0$ tangential anisotropy, and $\beta = 0$ isotropy.

\begin{figure}
\includegraphics[width=0.48\textwidth]{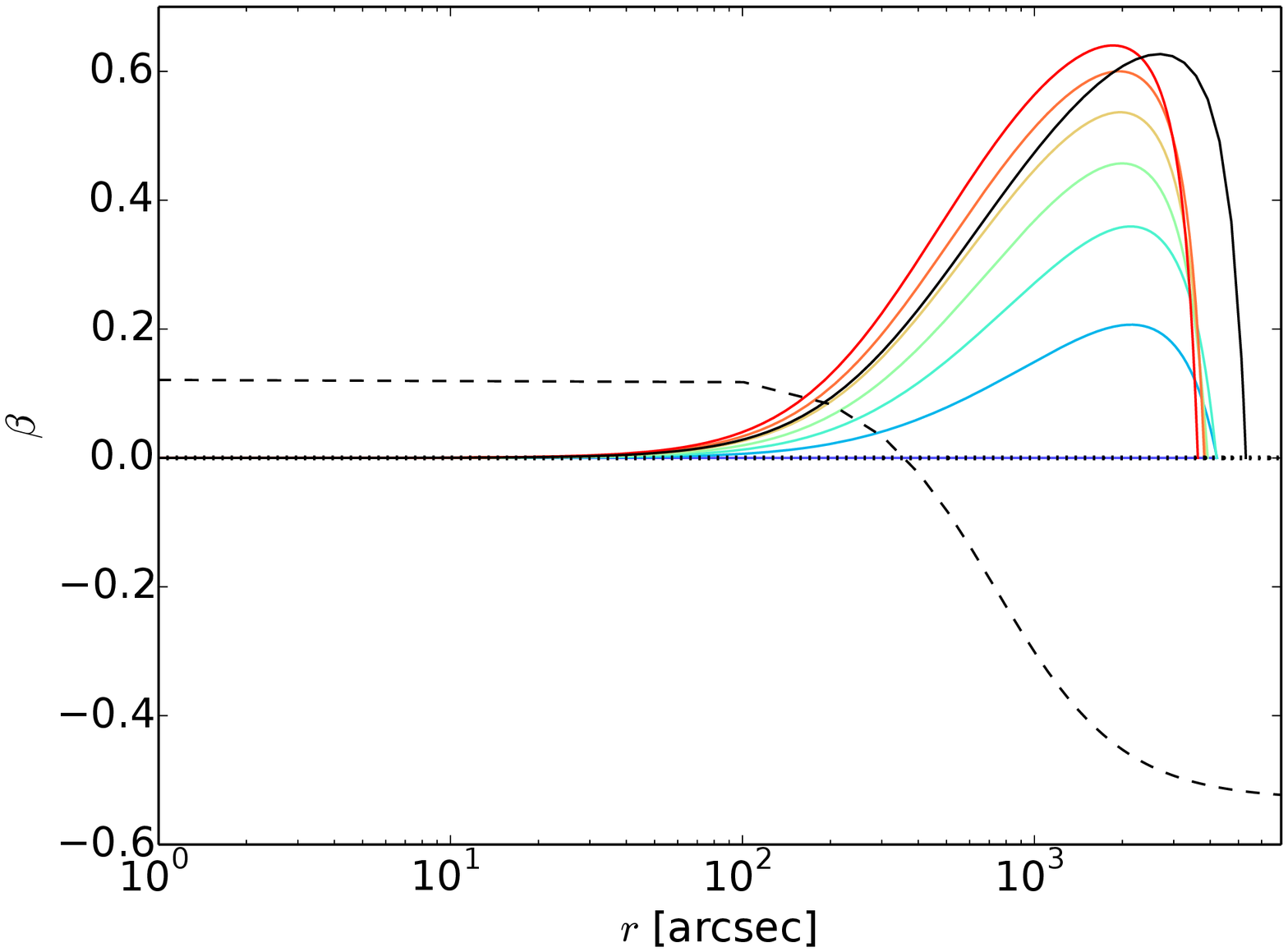}
\caption{Anisotropy radial profile for the models proposed as a fit to $\omega$ Cen. The solid lines are the radial profiles of the anisotropy for the best-fit \limepy\ models we obtained as a results of our fitting procedures, and are indicated with the same colours as in Fig.~\ref{Fig_SM_PM} and \ref{Fig_All_Kin}. The anisotropy profiles corresponding to the best-fit models proposed by \citet{Noyola2010} and \citet{vdMA2010} are indicated with the dotted and dashed lines, respectively.}
\label{Fig_cfr_beta}
\end{figure}

The radial profiles of the anisotropy for the best-fit \limepy\ models we obtained as a results of our fitting are indicated in Fig.~\ref{Fig_cfr_beta} with the same colours as in the other figures of this paper. These profiles are characterised by isotropy in the central part of the cluster, radial anisotropy in the intermediate part and isotropy again at the boundary. We notice that the different procedures followed in the fit introduce some differences in the anisotropy profiles: when the surface brightness profile is the only one used to determine the model parameters, the models are more compact and the peak of the anisotropy profile is located closer to the centre.

To better understand the role of anisotropy in the search for signatures from an IMBH, we compare the \limepy\ models, that do not account for the presence of an IMBH, to other models that provided a good fit for the kinematical profiles of $\omega$ Cen when a central IMBH was considered; we show the anisotropy profiles of these models in Fig.~\ref{Fig_cfr_beta}. The black dotted line is representing an isotropic profile: we show this to account for the isotropic models used in the analysis by \citet{Noyola2010}, so that the comparison with the \limepy\ models is more immediate. Moreover, we indicate the anisotropy profile obtained by \citet{vdMA2010} with the black dashed line: this profile describes a system which is radially anisotropic in the centre, and becomes tangentially anisotropic in the outer parts. 

We point out that all the models shown in Fig.~\ref{Fig_cfr_beta} are able to reproduce in a remarkable way the surface brightness profile of the cluster, despite their different dynamics. Some of them are also very accurately representing some of the kinematical profiles available for $\omega$ Cen. We reiterate that our two-steps fitting strategy mimics the analysis based on the Jeans equation, where the assumption on the presence and on the flavour of the anisotropy is only playing a role after the fit on the surface brightness has been carried out. 

The anisotropy plays an important role in shaping the relevant profiles for the cluster, but at the moment the available data are not enough to put a strong constraint on it. Indeed, the HST data in the centre of the cluster seem to indicate the presence of a larger amount of anisotropy than the one described by the models, while the ground-based data seem to indicate that less anisotropy is needed. More proper motions data, especially for stars located in the intermediate and external part of the cluster, will become available in the near future thanks to Gaia: these new data will ease the tension between the HST data and ground-based data and establish the amount of radial anisotropy in $\omega$ Cen.

\subsection{Comparison with previous works and implications on the presence of an IMBH}

\begin{figure}
\includegraphics[width=0.48\textwidth]{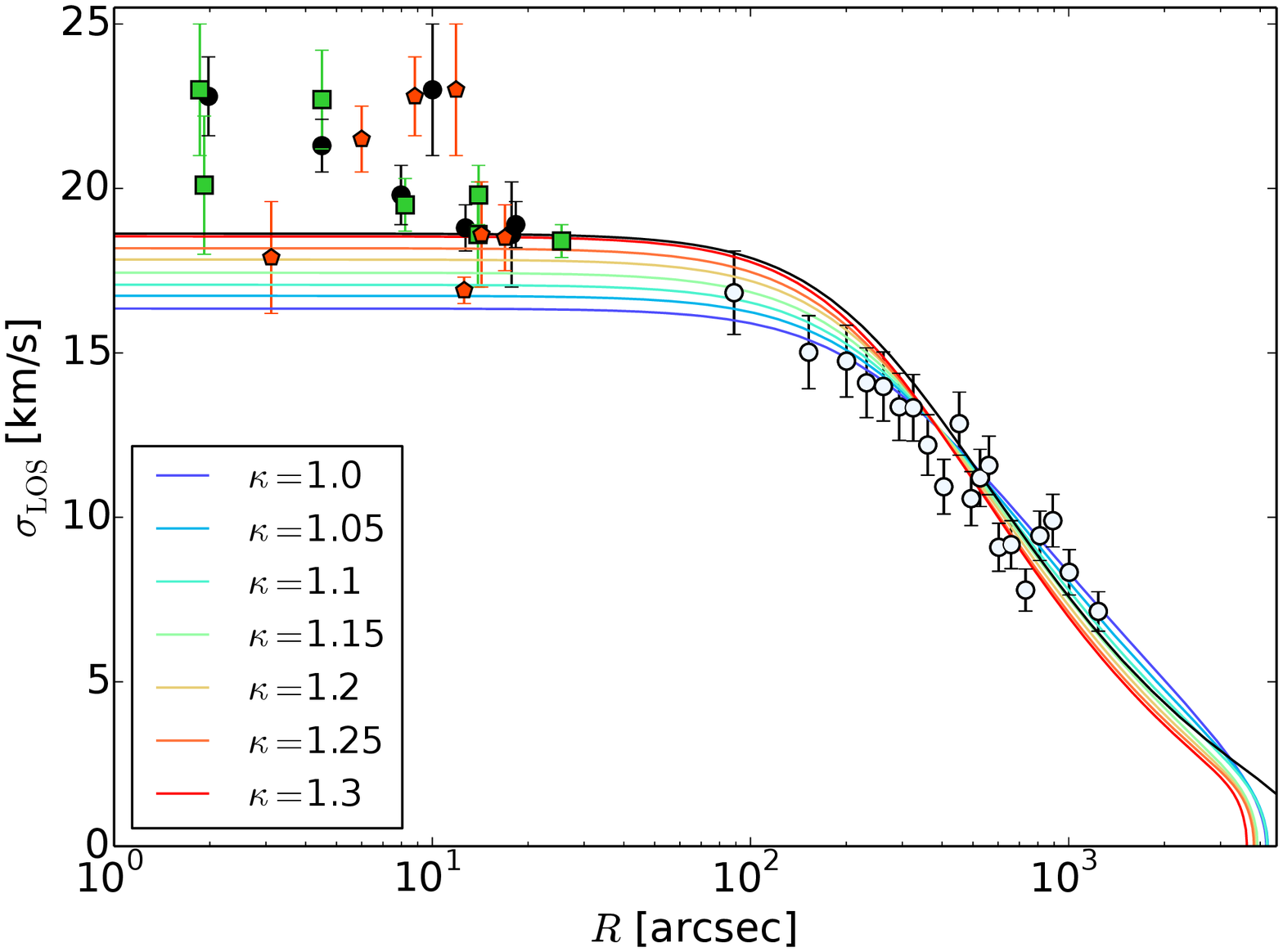}
\caption{Line-of-sight (projected) velocity dispersion profile of $\omega$ Cen, $\sigma_{\rm LOS}$. The black dots reproduce the profile provided by \citet{Noyola2010} when considering their kinematical centre, the green squares the profile obtained when considering the centre by \citet{Noyola2008}, and the red pentagons the one with the centre by \citet{AvdM2010}. Open circles reproduce data from \citet{Reijns2006} and \citet{Pancino2007}, with the profile calculated as in \citet{BVBZ2013}. Error bars are shown. The coloured lines reproduce the best-fit profiles obtained for the different choices of anisotropy content, as indicated by the labels; the black solid line represents the best-fit model obtained in our one-step fitting procedure.}
\label{Fig_cfr_VD}
\end{figure}

\subsubsection{IFU data to measure the central velocity dispersion}

It is particularly instructive to compare the results obtained here with the data that have been used to infer the presence of an IMBH in the centre of this cluster. In particular, we refer to the three velocity dispersion profiles provided by \citet{Noyola2010} and calculated from integrated spectra they collected in a region of the cluster very close to the centre. Because different estimates of the position of the centre of $\omega$ Cen have been obtained by using diverse techniques, different radial profiles can be calculated, by binning the data into annuli around the centres. \citet{Noyola2010} provided a first profile calculated by assuming the position of the centre they calculate based on their kinematic data, another profile by using the centre proposed by \citet{Noyola2008} and based on photometric data, and a third profile by considering the centre proposed by \citet{AvdM2010}, which is calculated from proper motions data. They point out that only in the case in which the centre is determined by analysing the kinematics of the cluster, it is possible to detect a cusp towards the centre in the velocity dispersion. The discrepancy in the profiles due to the choice of a different centre is only visible in the innermost part of the cluster (within $\sim$10 arcsec from the centre): the velocity dispersion profiles we calculated with the discrete line-of-sight velocity and proper motions data do not change significantly when considering a different centre.

Figure~\ref{Fig_cfr_VD} shows the line-of-sight velocity dispersion of $\omega$ Cen. The solid black line represents the best-fit \limepy\  model we obtained by fitting on all the data, while the coloured lines are the best-fitting profiles resulting from the two-steps analysis for different amounts of anisotropy, as indicated in the legend. Open circles reproduce data from \citet{Reijns2006} and \citet{Pancino2007}, with the profile calculated as in \citet{BVBZ2013}. The black dots reproduce the profile provided by \citet{Noyola2010} when considering their kinematical centre, the green squares the profile obtained when considering the centre by \citet{Noyola2008}, and the red pentagons the one with the centre by \citet{AvdM2010}. 

\begin{figure}
\includegraphics[width=0.48\textwidth]{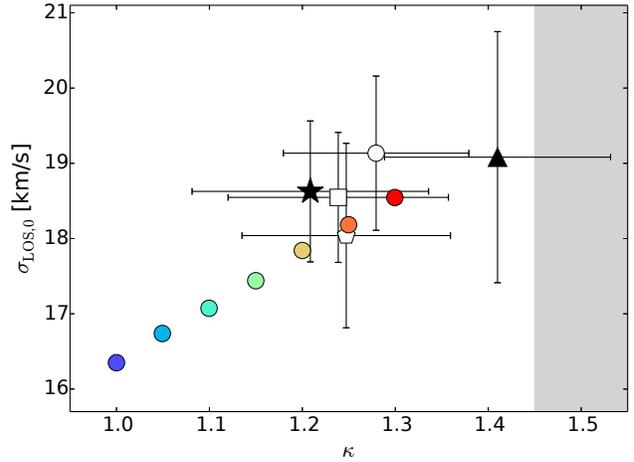}
\caption{Central velocity dispersion, $\sigma_{\rm LOS,0}$, as a function of $\kappa$, for the best-fit models considered here. The coloured dots reproduce the best-fit models obtained for the different choices of anisotropy content, with the same colours as the lines in Fig.~\ref{Fig_cfr_VD}; the black star represents the best-fit model obtained in our one-step fitting procedure. The black triangle shows the result obtained for a one-step fitting procedure carried out on the observational profiles when neglecting the ground-based proper motions from \citet{vanLeeuwen2000}. The white dot, white square, and white pentagon represent the result of a one-step fitting procedure carried out on the observational profiles with the addition of the profile provided by \citet{Noyola2010} when considering their kinematical centre, the profile obtained when considering the centre by \citet{Noyola2008}, and the one with the centre by \citet{AvdM2010}, respectively. The grey area on the plot shows the region of marginal radial orbit instability (we recall that the limit for the onset of instability proposed by \citealt{PolyachenkoShukhman1981} is $\kappa = 1.7 \pm 0.25$); 1-sigma uncertainties are shown for the results of one-step fitting procedures.}
\label{Fig_sigma0kappa}
\end{figure}

\begin{table*}
\begin{center}
\caption[Comparison with literature.]{Comparison with literature. We list the values of mass $M$, mass-to-light ratio $M/L$, and distance $d$ found in previous dynamical studies of $\omega$ Cen. References for each work are given in the first column, and a brief description of the models used is provided in the last column. Values between square brackets are not the result of the fitting procedure, but are fixed beforehand.} 
\label{Tab_Cfr_Literature}
\begin{tabular}{ccccccccccc}
\hline\hline
Reference  & $M$ & $M/L$ & $d$ & Models \\
  & [$10^6$ M$_{\odot}$] & [M$_{\odot}$/L$_{\odot}$] & [kpc] & \\
\hline
\citet{Meylan1987}   & 3.9                        & 2.9                     & [5.2]                   & multi-mass anisotropic \citet{Michie1963} models \\
\citet{Meylan1995}   & 5.1                        & 4.1                     & [5.2]                   & multi-mass anisotropic \citet{Michie1963} models \\
\citet{vandeVen2006} & 2.5 $\pm$ 0.3              & 2.5 $\pm$ 0.1           & 4.8 $\pm$ 0.3           & axisymmetric rotating orbit-based models \\
\citet{vdMA2010}     & 2.8                        & 2.62 $\pm$ 0.06         & 4.73 $\pm$ 0.0          & anisotropic models (Jeans) \\
\citet{Watkins2013}  &                            & 2.71 $\pm$ 0.05         & 4.59 $\pm$ 0.08         & anisotropic models (Jeans) \\
\citet{BVBZ2013}     & 1.953 $\pm$ 0.16           & 2.86 $\pm$ 0.14         & 4.11 $\pm$ 0.07         & rotating models \citep{VarriBertin2012} \\
\citet{Watkins2015}  & 3.452 $^{+0.145}_{-0.143}$ & 2.66 $\pm$ 0.04         & 5.19 $^{+0.07}_{-0.08}$ & isotropic models (Jeans) \\
\citet{DeVitaBZ2016} & 3.116                      & 2.87                    & [5.2]                   & anisotropic $f^{(\nu)}_{\rm T}$ models \\
\citet{Baumgardt2017}&  2.95 $\pm$ 0.02           & 2.54 $\pm$ 0.26         & 5.00 $\pm$ 0.05 & $N$-body simulations \\
\hline
this work            & 3.24 $^{+0.51}_{-0.47}$    & 2.92 $^{+0.36}_{-0.32}$ & 5.13 $\pm$ 0.25         & anisotropic \limepy\ models \\
\hline
\end{tabular}
\end{center}
\end{table*}

By inspecting the figure we see that, even though the best-fit models have been calculated without using the innermost velocity dispersion measurements, the most anisotropic ones can partially reproduce their behaviour. When considering the isotropic model, instead, there is no way of reaching the central values of the velocity dispersion, which are heavily underestimated. However, the most anisotropic models among those considered go through the lower sequence of points from the three profiles. The cusp noted in the black profile would imply a lower IMBH mass if considered on top of the anisotropic models instead of with an isotropic one. For this reason, it is important to accurately describe the global dynamics of a stellar system, when trying to look for IMBHs, even if they only have a significant and direct effect on the stars that surround them, because all the stars in the cluster play a role in shaping the quantities we observe. 

To further explore the effect of anisotropy on the line-of-sight velocity dispersion in the centre of the system, $\sigma_{\rm LOS,0}$, we compare the values we obtain for this quantity for the different best-fit models considered in this paper. Each point in Fig.~\ref{Fig_sigma0kappa} corresponds to a different model, for which the value of $\kappa$ is given in the abscissa and the value of $\sigma_{\rm LOS,0}$ on the ordinate. The coloured points correspond to the result of the two-steps fitting procedure, and are indicated here with the same colours used in the other figures of this paper; the black star corresponds to the result of the one-step fitting procedure. The other points indicate the values obtained when carrying out a one-step fitting procedure on all the observational profiles, with the addition of the innermost line-of-sight velocity dispersion points obtained by IFU data and shown in Fig.~\ref{Fig_cfr_VD}. The circle, square, and pentagon in Fig.~\ref{Fig_sigma0kappa} refer to fits that take into account respectively the profiles indicated with black circles, green squares, and red pentagons in Fig.~\ref{Fig_cfr_VD}. Figure~\ref{Fig_sigma0kappa} shows that the central velocity dispersion increases when considering more radially anisotropic models. For models with a $\kappa$ differing by 0.3, the corresponding difference in $\sigma_{\rm LOS,0}$ is about $\sim$3~km\,s$^{-1}$. The black triangle in the figure shows the result obtained when when neglecting the ground-based proper motions \citep{vanLeeuwen2000} from the one-step fitting procedure. The difference in the resulting value of $\kappa$ is due to the different trends observed between the ground-based and the HST proper motions (see top right panel of Fig.~\ref{Fig_All_Kin} for example). Gaia data will provide more accurate proper motions of stars in the outermost parts of this cluster \citep{Pancino2017}, and will make it possible to study the dynamics of $\omega$ Cen on its entire extent.

In order to provide more stringent constraints on the presence of an IMBH in $\omega$ Cen, it is not only important to use detailed dynamical models, taking into account the important ingredients that are responsible for the observed kinematics, but also to have accurate kinematical data covering the entire extent of the cluster at disposal. For the central crowded parts of clusters, often integral field spectroscopy is preferred to the measurement of the velocities of individual stars. However, these two different methods to measure the velocity dispersion at the centre of globular cluster seem to provide discrepant results. An example is the controversy around the cluster NGC 6388, for which \citet{Luetz2013} measured a central velocity dispersion of $\sim 23$~km\,s$^{-1}$ with integrated spectroscopy, while \citet{Lanzoni2013} only found a value of $\sim 13$~km\,s$^{-1}$ when calculating the dispersion from measures of velocities of single stars. This discrepancy causes the estimate of the mass of a central IMBH to go from $\sim 1.7 \times 10^4$ M$_{\odot}$ \citep{Luetz2013} to $\sim 2 \times 10^3$ M$_{\odot}$ \citep{Lanzoni2013} when considering the two different profiles. \citet{Bianchini2015} explored this issue by means of Monte Carlo cluster simulations, and concluded that the presence of a few bright stars could introduce a significant scatter in the velocity dispersion measurements when luminosity-weighted IFU observations are considered.

\subsubsection{Dynamical studies of $\omega$ Cen}

In addition to the ones mentioned above, other dynamical studies have been carried out to understand the dynamics of $\omega$ Cen. A compilation of the values of mass, mass-to-light ratio, and distance found in previous studies is given in Table~\ref{Tab_Cfr_Literature}. 

By considering multi-mass \citet{Michie1963} models, \citet{Meylan1987} found that the best-fit model to reproduce the density and line-of-sight velocity dispersion of this cluster is radially anisotropic, with an anisotropy profile of the same shape as the one of \limepy\ models, and suggests that this is related to the large half-mass relaxation time of this system. The best-fit mass and mass-to-light ratio obtained by \citet{Meylan1987}, $M = 3.9 \times 10^6$ M$_{\odot}$ and $M/L = 2.9$ M$_{\odot}$/L$_{\odot}$, are consistent with the ones we found in our one-step fitting procedure. 

\citet{vandeVen2006} considered Schwarzschild's orbit superposition method to model the dynamics of this cluster. Their best-fit model is close to isotropic in the innermost part and becomes tangentially anisotropic in the outermost part, and it has total mass $M = (2.5 \pm 0.3) \times 10^6$ M$_{\odot}$, mass-to-light ratio $M/L = 2.5 \pm 0.1$ M$_{\odot}$/L$_{\odot}$, and distance $d = 4.8 \pm 0.3$ kpc. These values are smaller than the ones we found, but consistent within $2\sigma$. \citet{Watkins2013} carried out discrete dynamical modelling of $\omega$ Cen, by considering models with a constant anisotropy profile. They found that the best-fit model is radially anisotropic with $\beta=0.10$; they also determine the mass-to-light ratio of the cluster to be $M/L = 2.71 \pm 0.05$ M$_{\odot}$/L$_{\odot}$, and the distance to be $4.59 \pm 0.08$ kpc. 

By using rotating models, \citet{BVBZ2013} found smaller values for the mass ($M = (1.953 \pm 0.16) \times 10^6$ M$_{\odot}$) and for the distance ($d = 4.11 \pm 0.07$ kpc) of the cluster with respect to the ones provided here, but their mass-to-light ratio ($M/L = 2.86 \pm 0.14$ M$_{\odot}$/L$_{\odot}$) is consistent with ours (this is because their smaller distance implies a smaller total luminosity for the cluster, and this combined with the smaller mass they find is producing a mass-to-light ratio similar to ours). 

The best-fit model we presented here showed that radial anisotropy plays an important role in $\omega$ Cen. However, our models do not account for the rotation, and describe a cluster with an homogeneous stellar population (constant $M/L$): relaxing these assumptions could provide an even better agreement with the observational profiles.

\section{Summary and conclusions}
\label{Sect_Conclusion}

Observational claims of IMBH detection are often based on analyses of the kinematics of stars, and in particular on the observed rise in the velocity dispersion profile towards the centre of the system \citep{vdMA2010,Noyola2010}. Here we considered the degeneracy between this IMBH signal and the presence of radially-biased pressure anisotropy in the system. As an example, we analysed the case of $\omega$ Cen, which has been at the centre of a controversy regarding the presence of an IMBH and its estimated mass.

In order to provide an analysis similar to the one based on the Jeans equations when a central IMBH is included, we considered a family of dynamical models, the \limepy\ models \citep{GielesZocchi}, and we compared them to the observational profiles of $\omega$ Cen by means of a two-steps fitting procedure. First, we fit the models with a fixed amount of anisotropy to the surface brightness profile of the cluster, to determine the model parameters, the total luminosity and the radial scale. Then, we determine the mass-to-light ratio of the cluster by fitting the vertical offset of the line-of-sight velocity dispersion. The results we get at the first step of this procedure show that models with different degrees of anisotropy are able to reproduce the surface brightness profile of $\omega$ Cen in great detail, and in a remarkably similar way. This strategy, therefore, allowed us to compare the kinematic profiles of models with a different dynamics that represent the surface brightness profile equally well. We found that anisotropic models are more adequate to reproduce both the line-of-sight and the proper motions velocity dispersion profiles of this cluster.

We also carried out an additional fit, by considering all the observational profiles at once, and without fixing a priori the amount of anisotropy to be expected in the system. We again found that a radially anisotropic model is providing the best-fit to the measured profiles.

Finally, we compared our best-fit models to the central line-of-sight velocity dispersion profile measured by means of integrated spectroscopy, which was used to claim the presence of an IMBH in the system \citep{Noyola2010}. We found that the most anisotropic models we considered are partially able to reproduce the innermost behaviour of the velocity dispersion profile, while the isotropic model predicts a much lower value for the central dispersion. The discrepancy between the models without IMBH and the data is smaller in the case of anisotropic models: if this were due to the presence of an IMBH, it would have a much smaller mass than predicted when considering isotropic models. This analysis cannot rule out the presence of an IMBH, but puts some caution on the amount of mass that could be accounted for by such an object.

The models we considered here were chosen specifically to explore the role of pressure anisotropy in shaping the projected profiles that can be compared with the observed ones. Indeed, pressure anisotropy plays an important role in the dynamics of globular clusters, and it should be taken into account to properly describe these systems. However, these models do not consider two ingredients that could be also important in determining these observational properties of clusters: rotation and mass segregation. In particular, as suggested also by \citet{Baumgardt2017}, the presence of a population of heavy remnants in the centre has the effect of increasing the line-of-sight velocity dispersion of visible stars towards the centre of the cluster \citep[for an example, see][]{Peuten2016}. We plan to explore the role of these dynamical ingredients in a follow-up study.

We reiterate that so far no strong evidence for the presence of an IMBH in the centre of a globular cluster has been found based on kinematic data. Upcoming measurements of proper motions by Gaia will soon enable us to study the kinematics of many Galactic globular clusters up to their outermost regions. This will unveil the kinematic properties of regions of the clusters that are so far unexplored, allowing us to obtain information on the rotation and on the velocity anisotropy of these stars, and on their interaction with the external tidal field. By studying the dynamics of these systems as a whole, we will be able to provide important information on the presence of IMBHs in their centres.

\section*{Acknowledgements}
The authors enjoyed fruitful discussions with Holger Baumgardt, Philip Breen, Ian Claydon, Alessia Gualandris, Douglas Heggie, Miklos Peuten, Justin Read, Anna Lisa Varri, and Laura Watkins. We also thank Antonio Sollima for comments on an earlier version of the manuscript, and the anonymous referee for constructive feedback. We are grateful to Dan Foreman-Mackey for providing the \emcee\ software and for maintaining the online documentation. AZ acknowledges financial support from the Royal Society (Newton International Fellowship), MG acknowledges financial support from the Royal Society (University Research Fellowship). MG and AZ acknowledge the European Research Council (ERC-StG-335936, CLUSTERS). VH-B acknowledges support from the Radboud Excellence Initiative Fellowship. 

\bibliographystyle{mn2e}
\bibliography{biblio.bib}


\end{document}